\documentclass[prl,twocolumn,superscriptaddress]{revtex4}
\usepackage{amsmath,graphicx}
\usepackage{epsfig}
\usepackage{paralist}
\usepackage{amsmath,amssymb,mathrsfs}
\usepackage{underscore}
\usepackage{natbib}
\usepackage{color}
\usepackage{dsfont}
\usepackage{bm}
\usepackage{ulem}
\usepackage[colorlinks,linkcolor=blue,anchorcolor=blue,urlcolor=blue,citecolor=blue]{hyperref}

\begin{document}

\title{Topological charge pumping in spin-dependent superlattices with glide symmetry}

\author{Qianqian Chen}
\author{Jianming Cai}
\author{Shaoliang Zhang}
\email{shaoliang@hust.edu.cn}
\affiliation{School of Physics, Huazhong University of Science and Technology, Wuhan 430074, China}
\affiliation{International Joint Laboratory on Quantum Sensing and Quantum Metrology, Huazhong University of Science and Technology, Wuhan, 430074, China}

\date{\today}

\begin{abstract}
Topological charge pumping represents an important quantum phenomenon that shows the fundamental connection to the topological properties of dynamical systems. Here, we introduce a pumping process in a spin-dependent double-well optical lattice with glide symmetry. In the dynamic process, the glide symmetry protects the band touching points and topological properties of the system {are} characterised by the non-Abelian Berry curvature. By engineering suitable form of interaction between different spin components, the model not only demonstrates topological phase transition, but also shows hybridisation between the spatial and temporal domain with novel topological features captured by the Wilson line along the synthetic directions. Our work {provides} a new model based on ultracold atoms towards the implementation of versatile topological matters and topological phenomena.
\end{abstract}

\maketitle

{\it Introduction.---} Topological charge pumping \cite{thouless}, also known as the dynamical quantum Hall effect, is a very robust transport process of particles through an adiabatic periodic evolution of the underlying Hamiltonian. In contrast to its classical counterpart, the transport in topological charge pumping is quantized and is not influenced by the perturbation. Such an important quantum phenomenon is connected to the topological feature of an effective two-dimensional (2D) system, in which the temporal domain induces a synthetic dimension that is perpendicular to the direction of {the lattice}. The pumped charge in one cycle is characterised by the Chern number of this 2D system \cite{polar1,polar2,liangfu, Citro}.
In condensed matter physics, the topological charge pumping is not easily observed because of the challenging requirement for flexible Hamiltonian engineering. Ultracold atom systems, with the highly controllable properties, provide a perfect platform to observe these topological phenomena \cite{Spielman1,Spielman2}. The charge pumping has already been observed experimentally in ultracold atom systems by driving a superlattice adiabatically \cite{cpump1,cpump2}. Furthermore, striking manifestation of topological charge pumping, such as spin pumping \cite{spump1} has also been observed in ultracold atom systems. Recently, topological charge pumping in an effective four-dimensional (4D) system \cite{4dpump1,4dpump2}, were realised in ultracold atom systems with the measurement of the corresponding second Chern number. Despite these exciting developments, the pumping processes in ultracold atom systems with nonsymmorphic symmetry remain largely unexplored yet.
Such pumping processes are expected to lead to novel quantum phenomena because of the interplay between nonsymmorphic symmetry and topological features.\cite{bandtouching1,bandtouching2,bandtouching3,bandtouching4,bandtouching5}.

%

%
In this work, we construct a one-dimensional (1D) spin-dependent superlattice with band touching points which are protected by the glide symmetry. By modulating the coupling amplitude between different spin components periodically in the pumping process, {the topological} phase transition can be observed. Although the total particle number pumped is still quantized based on the non-Abelian topological properties of this effective 2D system, the particle number of each spin component pumped in one adiabatic driving cycle is not necessary to be an integer anymore. More interestingly, with the appropriate form of coupling between different spin components, the system would exhibit glide symmetry along the hybridised direction of {the spatial} and temporal domain. The period of Wilson line along this direction is twice as much as the original dynamical system. The present proposal is experimentally feasible based on current state-of-art technology, and therefore provides a new platform to explore the intriguing topological phenomena with nonsymmorphic symmetry in ultracold atom systems.
{\it Model.---} The time-dependent Hamiltonian in the pumping process can be expressed as follows
\begin{equation}
\begin{split}
\mathcal{H}&(x,t)=\sum_\sigma\int dx\psi^\dag_\sigma(x)\Big\{-\frac{\hbar^2}{2m}\frac{\partial^2}{\partial x^2}-V_s\cos^2\Big(\frac{2\pi x}{d}\Big)+V_l\sigma_z \\
&\sin\Big(\frac{2\pi x}{d}+\varphi\Big)\Big\}\psi_\sigma(x)+\int dx\big\{B(\varphi)\psi^\dag_\uparrow(x)\psi_\downarrow(x)+h.c.\big\}
\label{dssh1}
\end{split}
\end{equation}
where $\sigma=\uparrow,\downarrow$ characterise different hyperfine spin states and $\sigma_z=\pm 1$. $V_s$ and $V_l$ are the lattice depth for lattices with short and long lattice spacing respectively. The short lattice with $\lambda_s=d/2$ is spin independent, but the long lattice with $\lambda_l=d$ is spin dependent which means different spin components feel opposite lattice potentials \cite{Finkelstein,Deutsch,Haycock,Mandel,DMckay,PSoltan,BYang}. $B(\varphi)$ is the microwave field coupling different spin components, the amplitude and phase can be well controlled in experiment. The time-dependent parameter $\varphi(t)=\varphi_0+\omega t$ describes the moving of spin-dependent lattice along the $x$ direction, where $\omega$ is the frequency of the pumping process which was chosen to make sure the evolution is adiabatic. The evolution of this lattice potential is shown in Fig. 1(a).

\begin{figure}
\begin{center}
{\includegraphics[width=0.48 \textwidth]{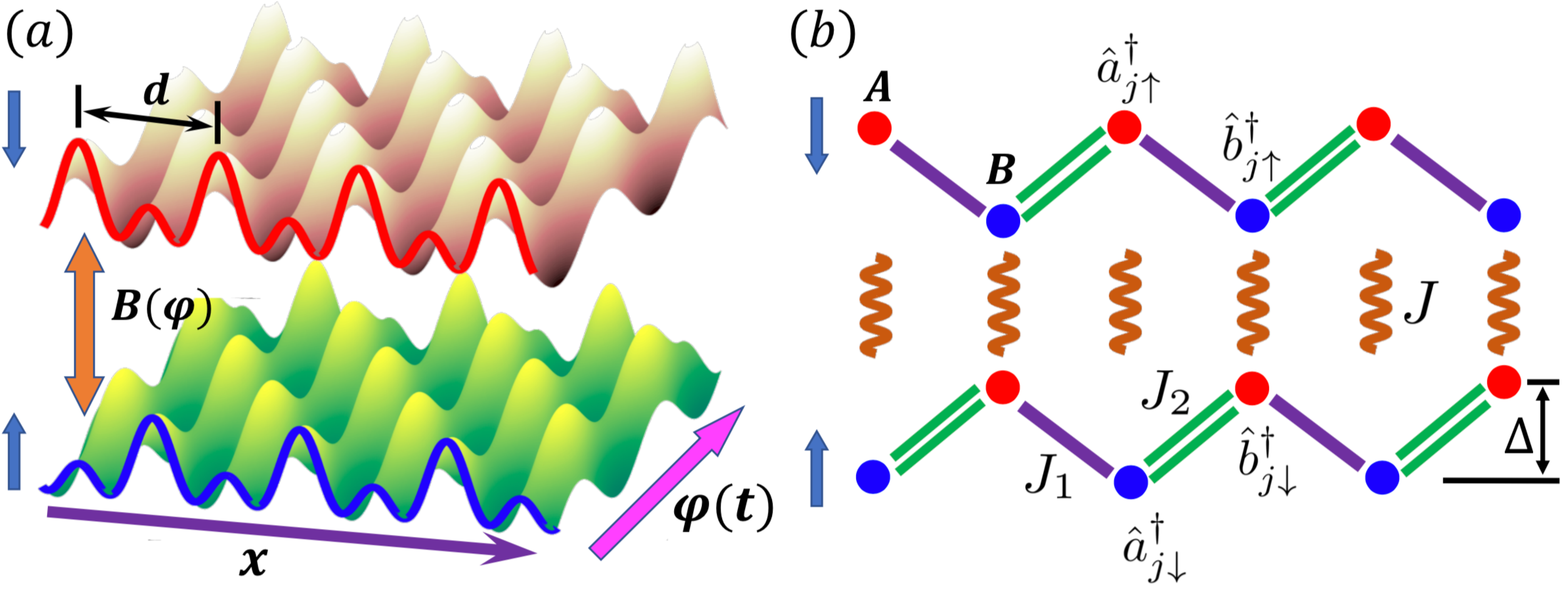}}
\caption{ (a) The evolution of lattice potentials with different spin components. The lattice potentials of spin-up and down are shown by brown and green curves respectively. The red and blue lines along $x$ direction (purple arrow) correspond to the lattice potential at $\varphi=0$ for spin-up and down respectively. Lattice potentials of different spin components are driven with the same way along $\varphi(t)$ direction (magenta arrow). A microwave field $B(\varphi)$ provides the coupling between spin-up and down.  (b) Tight-binding model of the one-dimensional superlattice for an arbitrary $\varphi$. Red and blue dots represent different sublattice sites. The intra-well(inter-well) tunneling $J_1$($J_2$) is shown by single purple(double green) line. The on-site coupling $J$ is shown by orange spiral line. $\Delta$ is the energy detuning between different sublattices.}
\end{center}
\end{figure}

In the extreme case $B(\varphi)=0$, the system can be considered as two independent superlattices. For each spin component, the driving process is equivalent to Thouless pumping which was already observed in ultracold atom experiments \cite{cpump1,cpump2}. We can explain this equivalence briefly using the tight-binding Hamiltonian as
\begin{equation}
\begin{split}
\mathcal{H}(\varphi)=&-\sum_n\big(J_1(\varphi)a^\dag_nb_n+J_2(\varphi)b^\dag_na_{n+1}+h.c.\big) \\
&+\frac{\Delta(\varphi)}{2}\sum_n\big(a^\dag_na_n-b^\dag_nb_n\big)
\label{pump1}
\end{split}
\end{equation}
where $J_1(\varphi)$ and $J_2(\varphi)$ are the nearest neighbor tunneling, $\Delta(\varphi)$ is the energy detuning between $A$ and $B$ sublattices, as one of two chains shown in Fig. 1(b) and all these parameters change periodically with $\varphi$. In the adiabatic process, the temporal domain can be regarded as a synthetic dimension perpendicular to {the spatial} domain and $\varphi$ as the corresponding quasi-momentum, then the time-dependent Hamiltonian (\ref{pump1}) can be considered as an effective 2D static Hamiltonian. The Berry curvature in the first Brillouin Zone (BZ) and Chern number can be well defined and one can easily find that the Chern number of the lowest band is $\mathcal{C}=1$ \cite{LWang}.

If $B(\varphi)\ne 0$, different spin components couple with each other and different quantum phenomena arise. In this case, the total Hamiltonian (\ref{dssh1}) including two different spin components can also be expressed using tight-binding approximation as
\begin{equation}
\begin{split}
\mathcal{H}&=J_1(\varphi)\sum_n\big(a^\dag_{n,\uparrow}b_{n,\uparrow}+b^\dag_{n,\downarrow}a_{n+1,\downarrow}\big)+J_2(\varphi)\sum_n\big(b^\dag_{n,\uparrow}a_{n+1,\uparrow} \\
&+a^\dag_{n,\downarrow}b_{n,\downarrow}\big)+J(\varphi)\sum_n\big(a^\dag_{n,\uparrow}a_{n,\downarrow}+b^\dag_{n,\uparrow}b_{n,\downarrow}\big)+h.c. \\
&+\frac{\Delta(\varphi)}{2}\sum_n\big(a^\dag_{n,\uparrow}a_{n,\uparrow}-b^\dag_{n,\uparrow}b_{n,\uparrow}-a^\dag_{n,\downarrow}a_{n,\downarrow}+b^\dag_{n,\downarrow}b_{n,\downarrow}\big)
\label{tbdssh1}
\end{split}
\end{equation}
where $J(\varphi)$ is the coupling between different spin components which is proportional to $B(\varphi)$. One can also calculate the band structure and other physical properties of this effective 2D system.

With any form of the coupling strength $J(\varphi)$, the band touching point emerges at the edge of the BZ where $k=\pm\pi/d$ for an arbitrary $\varphi$, as shown in Fig. 2. Because of the degeneracy, the non-Abelian Berry curvature is required to describe the topological properties of this effective 2D system. The matrix form of the non-Abelian Berry curvature is \cite{Culcer,Chang1}
\begin{equation}
\boldsymbol{\mathcal{F}}=\partial_k\boldsymbol{\mathcal{A}}_\varphi-\partial_\varphi\boldsymbol{\mathcal{A}}_k-i[\boldsymbol{\mathcal{A}}_k, \boldsymbol{\mathcal{A}}_\varphi]
\end{equation}
where $\boldsymbol{\mathcal{A}}_k=i\langle{\bf u}|(\partial/\partial k)|{\bf u}\rangle$ and $\boldsymbol{\mathcal{A}}_\varphi=i\langle{\bf u}|(\partial/\partial\varphi)|{\bf u}\rangle$ are Berry-Wilzeck-Zee connections \cite{Berry,Wilzeck} which are both $2\times 2$ matrix where $|{\bf u}\rangle=\{|u_n(k,\varphi)\rangle\}$ $(n\le 2)$ is the instantaneous periodic Bloch wave function of the lowest two bands at an arbitrary $\varphi$ (where $|\psi_n(k,\varphi)\rangle=e^{ikx}|u_n(k,\varphi)\rangle$ is the corresponding Bloch wave function). The particle number transported within one cycle can be characterised with the non-Abelian Chern number \cite{Chang2,Sundaram,dixiao}
\begin{equation}
Q=-\frac{1}{2\pi}\int^{2\pi}_0\int_\mathrm{BZ}\mathrm{tr}(\boldsymbol{\mathcal{F}})dkd\varphi
\label{napump}
\end{equation}
It means that the pumped charges in each cycle should also be quantized. We should point out this conclusion only relevant when the driving frequency $\omega$ satisfying the condition $\mathcal{W}\ll\hbar\omega\ll\mathcal{D}$, where $\mathcal{W}$ is the band width of lowest two bands and $\mathcal{D}$ is energy gap between lowest two bands and the other upper bands. Only in this case, the lowest two bands can be approximated as degenerate in the whole BZ and the probability of the excitation to the upper bands can be ignored, which means the dynamic process can be considered as an adiabatic evolution in the subspace of lowest two almost degenerate bands.

\begin{figure}
\begin{center}
{\includegraphics[width=0.48 \textwidth]{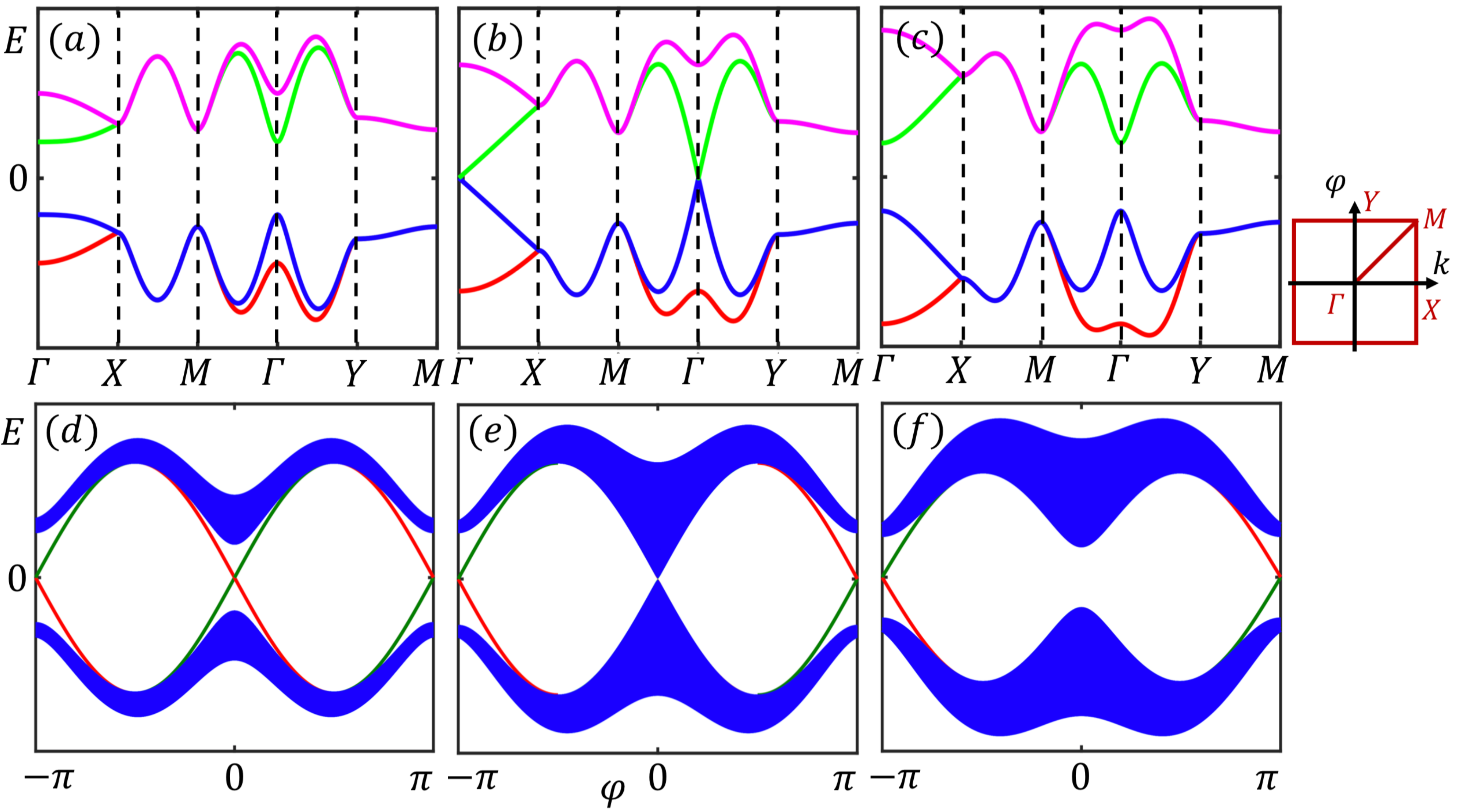}}
\caption{(a-c). Band structure of the effective 2D system in different topological phases, where the coupling has the form as $J(\varphi)=J(1+\cos\varphi)$. (a) $J<(J_1+J_2)/2$, (b) $J=(J_1+J_2)/2$, (c) $J>(J_1+J_2)/2$. (d-f) The corresponding energy spectrum for different $\varphi$ in one-dimensional superlattice with open boundary condition. (d) When $J<(J_1+J_2)/2$, there are two zero energy modes at $\varphi=0$ and $\pi$ respectively. (e) When $J=(J_1+J_2)/2$, there is a band touching point at $\varphi=0$ and the topological phase transition occurs. (f) When $J>(J_1+J_2)/2$, the band gap reopens and there is only one zero energy mode at $\varphi=\pi$. }
\end{center}
\end{figure}

{\it Topological phase transition.---} In the extreme case $J(\varphi)=0$, the lowest two bands are degenerate in the whole BZ and the Chern number of the effective 2D system should be $C=2$.
With the increase of the amplitude of {the microwave field}, the gap will be opened in most of the BZ. But at $k=\pm\pi/d$, the degeneracy will be protected by glide symmetry \cite{dssh}. With different forms of coupling $J(\varphi)$ corresponding to different types of glide symmetry, different quantum phenomena can be observed. So the form of $J(\varphi)$ is crucial for the non-Abelian topological properties in this effective 2D system. In the following, we will discuss two typical types of coupling and {their} topological consequences.

{\it Case 1:} $J(\varphi)$ is real. In this case, with the combination of a spatial translation of half of the lattice spacing $d/2$ and the spin flip $\uparrow\leftrightarrow\downarrow$, the Hamiltonian (\ref{dssh1}) is conserved. This operation is independent of $\varphi$ which means the spatial and temporal domain is separated, we can call the system has {\it spatial glide symmetry}.
If $J(\varphi)=2J$ is a constant, with the increase of the amplitude of {the microwave} field, the topological phase transition emerges at $2J=J_1+J_2$ and the Chern number of lowest two bands becomes $C=0$ when passing across this critical point.
But if the amplitude of $J(\varphi)$ has a modulation in temporal domain, for example $J(\varphi)=J(1+\cos\varphi)$ where $J>0$. With the increase of the amplitude, the topological phase transition emerges at the same point but the Chern number of lowest two bands will {become} $C=1$ instead of $C=0$.
 In {the experiment}, the topological phase transition can be directly observed by measuring the movement of center-of-mass (COM) of the atoms in one cycle. One can prepare the initial state in the Mott-Insulator phase with the lowest two bands full filled. Then the movement of COM should have a sharp transition from one side of the critical point to the other. It means that if the amplitude of coupling is modulated periodically, although the total particle number transported in one cycle is quantized to 1, each spin component transported can have no quantization property in {the pumping} process.


This topological phase transition can be explained as {follows}. With an arbitrary $\varphi$, the topological properties of the instantaneous 1D Hamiltonian are characterised by {the Zak} phase. Only two special times $\varphi=0$ and $\varphi=\pi$ need to be considered, because at these moments, this 1D system is a two-leg Su-Schrieffer-Heeger (SSH) model and the Zak phase should be $0$ or $\pi$ which depending on the coupling strength $J(\varphi)$, as discussed in our earlier work \cite{dssh}.
Because $J(\varphi)=J(1+\cos\varphi)$ has a modulation in temporal domain, the system only touches the critical point at $\varphi=0$ when $2J=J_1+J_2$. as shown in Fig. 2(a-c). If the system has open boundary condition in real space, the topological phase transition can also be characterised by the change of edge states, as shown in Fig. 2(d-f). Although the conclusions come from the tight-binding model, the numerical calculation with plane-wave expansion can get the same results, which was shown in supplementary materials \cite{supple}.

{\it Case 2:} $J(\varphi)$ is complex, namely $J(\varphi)=J(1+e^{i\varphi})$. In this case, the glide operation which conserves the Hamiltonian (\ref{dssh1}) is the combination of a spatial translation of half of the lattice spacing $d/2$ and the spin flip with an $\varphi$ dependent phase $\psi_\uparrow(x)\rightarrow\psi_\downarrow(x)$, $\psi_\downarrow(x)\rightarrow e^{-i\varphi}\psi_\uparrow(x)$. In this case, the spatial and temporal operation couple with each other leading to the fact that the spatial and temporal domain is unseparable \cite{cjwu}.
Therefore, the system demonstrates a novel type of symmetry, namely  {\it synthetic glide symmetry}.
With the increase of the coupling strength, the system also exhibits a topological phase transition from $C=2$ to $C=1$ at the critical point $2J=J_1+J_2$. It can be seen that the position of critical point just depends on the amplitude of the coupling $|J(\varphi)|=J\sqrt{2(1+\cos\varphi)}$.


{\it Wilson line.---} It looks like the above two cases have similar behaviors in topological phase transition. But we should point out that the topological properties of these two systems are different because of two different types of glide symmetry. In these effective 2D systems, {the Wilson line} is required to characterise the non-Abelian topological properties \cite{Wilzeck,Soluyanov}. For example, the Wilson line starting from $\Gamma=(0,0)$ point to an arbitrary point ${\bf k}=(k,\varphi)$ can be defined as
\begin{equation}
\boldsymbol{\mathcal{W}}_{\Gamma\rightarrow{\bf k}}=\mathscr{P}\exp\Big\{i\int_L\boldsymbol{\mathcal{A}}({\bf k})\cdot d{\bf k}\Big\}
\end{equation}
which is a path-ordered($\mathscr{P}$) integral and $L$ is path from $\Gamma$ to ${\bf k}$ in reciprocal space. In experiment, if the initial state was prepared on the eigenstate of the $n$-th band as $|\psi_n(0,0)\rangle$ with quasi-momentum $k=0$, the element of Wilson line can be estimated by detecting the population of atoms on the $m$-th band at quasi-momentum ${\bf k}$ after the dynamic evolution along route $L$ as $|W^{mn}_{\Gamma\rightarrow{\bf k}}|^2$ where
\begin{equation}
W^{mn}_{\Gamma\rightarrow{\bf k}}=\langle u_m(k,\varphi)|\boldsymbol{\mathcal{W}}_{\Gamma\rightarrow{\bf k}}|u_n(0,0)\rangle
\label{wilsonline}
\end{equation}
To ensure the above relation Eq. (\ref{wilsonline}) is correct, the criterion $\mathcal{W}\ll\hbar\omega\ll\mathcal{D}$ should also be satisfied due to the same reason in the discussion about Eq. (\ref{napump}). The Wilson line can be expressed as a sequence of path-ordered products of projectors and calculated numerically \cite{wilsonloop1,wilsonloop2,wilsonloop3}.

In the spatial-temporal separable system, because $\langle u_1(k,\varphi)|\partial_\varphi|u_2(k,\varphi)\rangle=0$, which means the eigenstates of the lowest two bands can not couple with each other in pumping process, the Wilson line is a trivial straight line. But in the spatial-temporal {inseparable system}, because of the hybridisation of the spatial and temporal domains, $\langle u_1(k,\varphi)|\partial_\varphi|u_2(k,\varphi)\rangle\ne 0$ and the tunneling between different bands will be induced in {the pumping process}. The form of {the Wilson line} will always be complex and depends on the route of adiabatic evolution.
In Fig. 3(b), the route of evolution is only in {the temporal domain} and the quasi-momentum is $k=0$, as the magenta arrow in Fig. 3(a), the population on the ground state decreases smoothly from $1$ but increases suddenly after going through the edge of BZ. The Wilson line has the same period as the original Hamiltonian in {the temporal domain}. In Fig. 3(c), we add a very small force to accelerate the atoms in {the spatial domain} and drive the system in {the temporal domain} simultaneously. By {fine-tuning} the force as $F=\hbar\dot{k}=\hbar\omega/d$,  one can control the atoms moving along the diagonal line of the BZ of this effective 2D system, as the green arrow in Fig. 3(a). After one period of driving, the state flips to the eigenstate of the first excited band which is orthogonal to the initial state. Only after another period, it can return to the initial one, which corresponds to a M\"{o}bius strip.
\begin{figure}
\begin{center}
{\includegraphics[width=0.48 \textwidth]{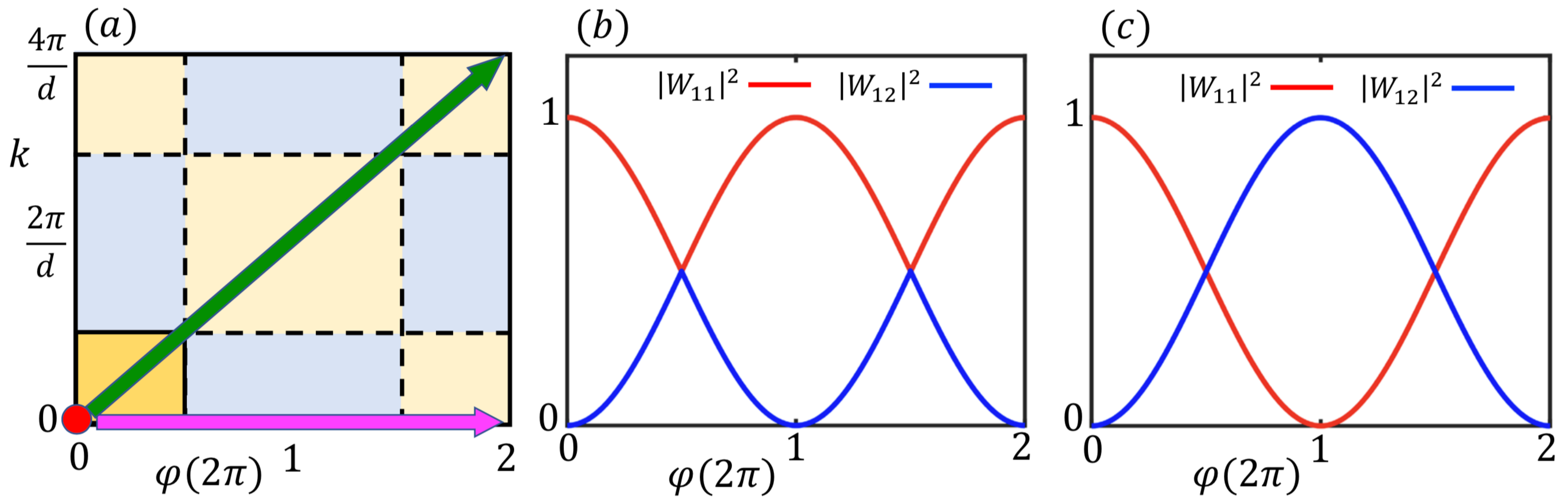}}
\caption{Wilson line and spin evolution. The initial state was prepared on the ground state at $k=0$. (a) The magenta and green arrows correspond to two pumping processes along different directions respectively. The yellow area is part of the 1st BZ. The light blue and light yellow areas correspond to the ground states that have ``$+$" and ``$-$" eigenvalues of {the glide operator} respectively. (b,c) The population on different bands in {the pumping process} along different directions. (b) is along the $\varphi$ direction and keeps $k=0$, (c) is along the diagonal line of the first BZ as $\varphi=kd$. The parameters are $V_s=8E_R$, $V_l=4E_R$ and $B_0=0.01E_R$.
}
\end{center}
\end{figure}

All these results arise from the synthetic glide symmetry. In the extreme case that the coupling amplitude approaches to zero, the lowest two bands are almost degenerate in the whole BZ and the Wilson line can be estimated analytically. If the pumping is along $\varphi$ direction as {the} magenta arrow shows in Fig. 3(a), the Wilson line has the matrix form as
\begin{equation}
W_{0\rightarrow\varphi}=e^{i\theta(\varphi)}\left(\begin{array}{cc} \cos(\varphi/4) & -i\sin(\varphi/4) \\ -i\sin(\varphi/4) & \cos(\varphi/4) \end{array}\right)
\label{wline}
\end{equation}
where the bases are the lowest two eigenstates of instantaneous Hamiltonian $|u^-_{k\varphi}(x)\rangle$ and $|u^+_{k\varphi}(x)\rangle$, and $\pm$ correspond to different eigenvalues of glide symmetry operator respectively. The $e^{i\theta(\varphi)}$ is the U(1) part with a complex form, the SU(2) part shows that the populations on ``-" and ``+" bands in evolution should be $|W_{--}|^2=\cos^2(\varphi/4)=[1+\cos(\varphi/2)]/2$ and $|W_{-+}|^2=\cos^2(\varphi/4)=[1-\cos(\varphi/2)]/2$ which have a period of $4\pi$ in temporal domain if the initial state is prepared on the ground state $|u^-_{k\varphi}(x)\rangle$. But in this pumping process, there is a band crossing point at the edge of BZ where $\varphi=\pi$ because of the synthetic glide symmetry, and after passing this point, $|u^+_{k\varphi}(x)\rangle$ becomes the ground state. So the population on the lowest band $|W_{11}|^2$ should flip from $|W_{--}|^2$ to $|W_{-+}|^2$, it will flip again after passing another edge. So the period of the population evolution is $2\pi$ as shown in Fig. 3(b). If the pumping is along {the} diagonal direction of BZ as green arrow shows in Fig. 3(a), the matrix form of {the} Wilson line is similar to Eq. (\ref{wline}), except that $\varphi$ is replaced by the parameter $p=(kd+\varphi)/2$ along the hybridised direction, but $|u^-_{k\varphi}(x)\rangle$ is always the ground state along this direction because of the synthetic glide symmetry. Then the population on the lowest band should always be $|W_{--}|^2$ and the period of evolution is $4\pi$ as shown in Fig. 3(c).
We point out that all the evolution of the population on the ground state can be directly measured in experiment if the initial state prepared in the superfluid (SF) phase instead of the MI phase. More analytical proof and detailed {discussions are} shown in supplementary materials \cite{supple}.


{\it Conclusion and discussion.---} In this work, we propose a charge pumping process by periodically driving a spin-dependent superlattice. Because of the glide symmetry, the band structure has nodal lines and the topological properties of this system should be characterised by non-abelian Berry curvature. In this system, we find that the topological phase transition can be well controlled by the form of coupling between different spin components. Especially, we can find the case that each spin component transported in one cycle is not quantized in our system, {although the} total particle number pumped is still quantized. In addition, the interplay between the glide symmetry and the coupling between different spin components can hybridise the temporal and spatial domains can induce the synthetic glide symmetry, which opens a route to novel topological phenomena, such as a M\"{o}bius strip without needing the twisted boundary. The spin-dependent superlattice system can be highly controlled experimentally and exhibit versatile physics. The generalisation of the present model allows {realising} the Floquet topological systems with nonsymmorphic symmetry using periodically driving process with high frequency \cite{cjwu, luning, hcpo, yangpeng, zjsu}. The influence of interaction on topological pumping processes may open up more interesting quantum phenomena.

{\it Acknowledgement} --- We thank the useful discussion with Xiang-Fa Zhou and S. L. Zhang also thanks Qi Zhou and Zheng-Wei Zhou for valuable suggestions. S. L. Zhang was supported by starting up funding of Huazhong University of Science and Technology. J. M. Cai was supported by National Key R\&D Program of China (Grant No. 2018YFA0306600).

\onecolumngrid

\begin{appendix}

\section{The result of topological property of the charge pumping process with exact calculation using plane-wave expansion}

Using the plane-wave expansion, we checked all results in {the} main text and find {consistent results}. We chose two types of inter-spin coupling where the spatial and temporal {domains} are separable and {inseparable} respectively.

\begin{figure}[tph]
\begin{center}
{\includegraphics[width=0.9 \textwidth]{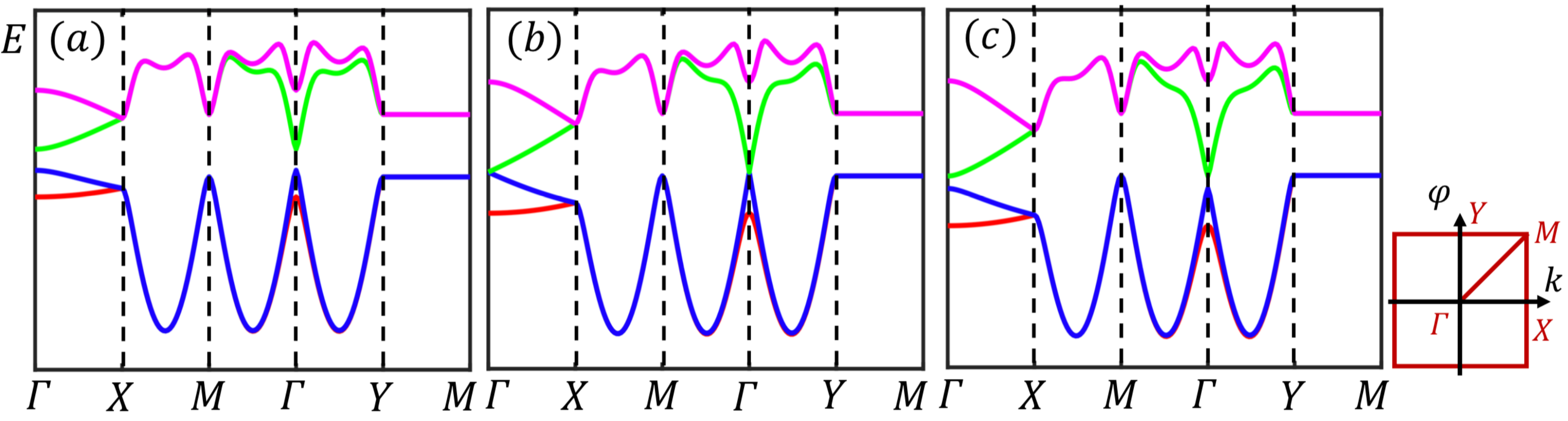}}
\caption{Band structure of the effective two-dimensional system with coupling $B(\varphi)=B_0(1+\cos\varphi)$ using plane-wave expansion. The lattice depths with short and long wave length are $V_s=V_l=6E_R$ respectively. (a) $B_0=0.5E_R$, (b) $B_0=0.8E_R$, (c) $B_0=1.0E_R$.}
\end{center}
\end{figure}

Consider the spatial-temporal separable system firstly. To generalize our results, the microwave field coupling different spin components should have the form $B(\varphi)=B_0(1+\cos\varphi)$ The band structure is shown in Fig. S1. One can see that the band structure of this effective 2D system using plane-wave expansion is similar to that deduced from tight-binding model in main text with only quantitative differences. The asymmetry of lower and upper two bands mainly comes from the next-nearest neighbor tunneling and the influence of the higher bands. Here we choose the lattice depth with short and long wave length to be $V_s=V_l=6E_R$ where $E_R=\frac{h^2}{2md^2}$ is the recoil energy. One can easily check that the topological properties are similar to the result of tight-binding model. When $B_0\approx 0.8E_R$, this effective 2D system has a band touching point at $\Gamma$ point thus has a topological phase transition, as shown in Fig. S1(b). Fig. S2 shows the trace of non-Abelian Berry curvature on two sides of this topological phase transition.

\begin{figure}[tph]
\begin{center}
{\includegraphics[width=0.5 \textwidth]{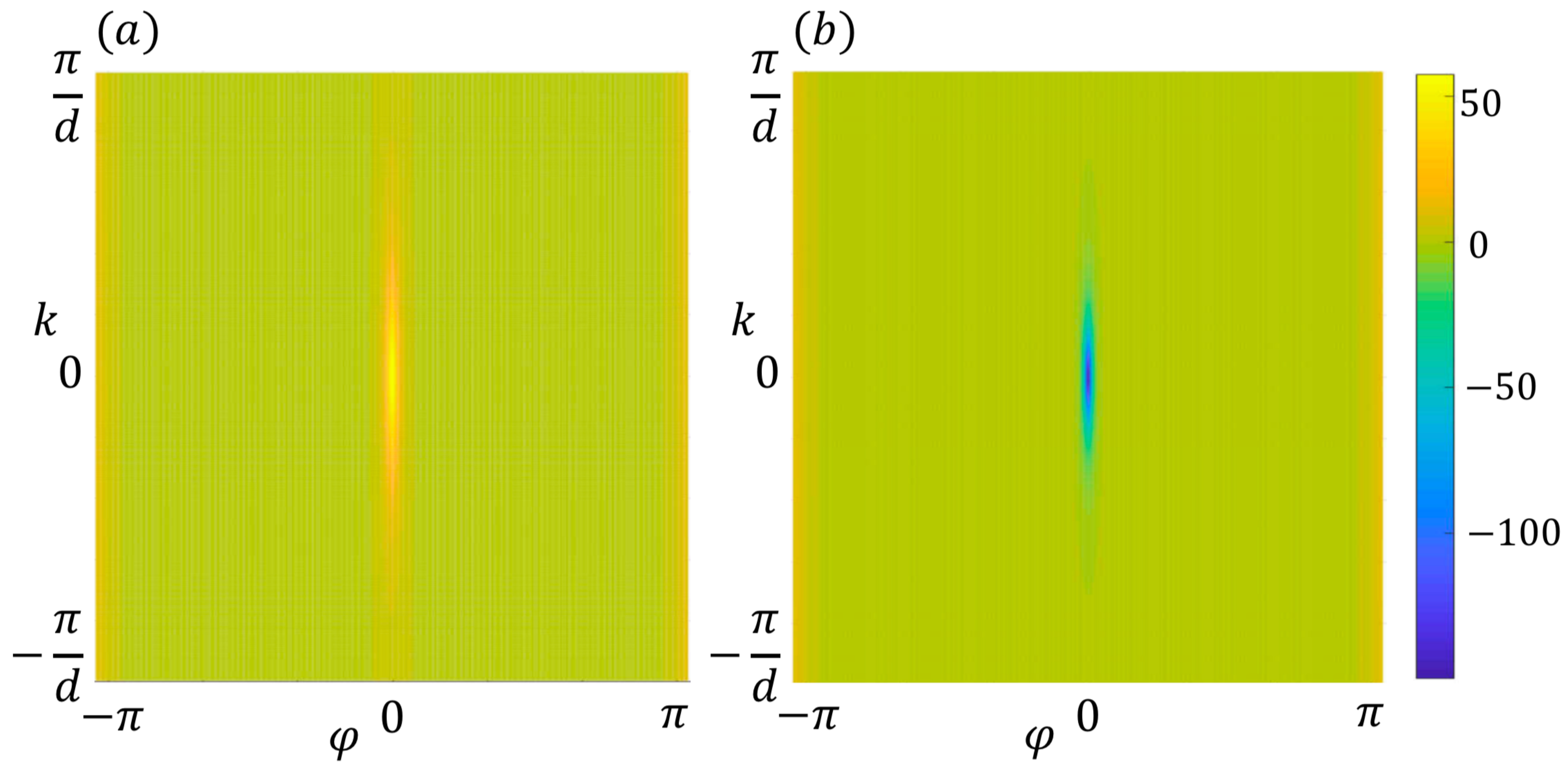}}
\caption{The trace of non-abelian Berry curvature of the effective two-dimensional system in different topological phases. The lattice depth with short and long wave length are $V_s=V_l=6E_R$ respectively. (a) $B_0=0.5E_R$, (b) $B_0=1.0E_R$.}
\end{center}
\end{figure}

Then we consider the spatial-temporal {inseparable} system. In this case, the microwave field coupling different spin components should have the form $B(\varphi)=B_0(1+e^{i\varphi})$ to preserve the synthetic glide symmetry. The band structure is almost same as Fig. S1 and the topological phase transition can also be observed.

\section{Discussion about the Wilson line}

Consider the extreme case $B(\varphi)=0$ first. The Hamiltonian of each spin component is
\begin{equation}
\mathcal{H}_\sigma(x,\varphi)=\int dx\psi^\dag_\sigma(x)\Big\{-\frac{\hbar^2}{2m}\frac{\partial^2}{\partial x^2}-V_s\cos^2\Big(\frac{2\pi x}{d}\Big)+V_l\sigma_z\sin\Big(\frac{2\pi x}{d}+\varphi\Big)\Big\}\psi_\sigma(x)
\end{equation}
which has the relation as $\mathcal{H}_\uparrow(x,\varphi)=\mathcal{H}_\downarrow(x+d/2,\varphi)$. Assume that the plane-wave solution of Bloch wave function on the lowest bands of instantaneous Hamiltonian $\mathcal{H}_\uparrow(x,\varphi)$ is
\begin{equation}
|\psi_{k\varphi,\uparrow}(x)\rangle=e^{ikx}|u_{k\varphi,\uparrow}(x)\rangle=\frac{e^{ikx}}{\sqrt{d}}\sum_\ell c_\ell(k,\varphi)\exp\Big\{\frac{i2\pi\ell x}{d}\Big\}|\uparrow\rangle
\end{equation}
which has the eigenenergy $\epsilon(k,\varphi)$. We have the relation $\sum_\ell |c_\ell(k,\varphi)|^2=1$ because of the normalization of the Bloch wave function. Then the corresponding plane-wave solution of instantaneous Hamiltonian $\mathcal{H}_\downarrow(x,\varphi)$ with the same eigenenergy should have the form as
\begin{equation}
|\psi_{k\varphi,\downarrow}(x)\rangle=e^{ikx}|u_{k\varphi,\downarrow}(x)\rangle=\frac{e^{ikx}}{\sqrt{d}}\sum_\ell (-1)^\ell c_\ell(k,\varphi)\exp\Big\{\frac{i2\pi\ell x}{d}\Big\}|\downarrow\rangle
\end{equation}
which satisfy the relation $\psi_{k\varphi,\uparrow}(x)=\psi_{k\varphi,\downarrow}(x+d/2)$. Then we consider the case that $B(\varphi)$ is not zero, different spin {components} and different energy bands with the same quasi-momentum $k$ will be coupled and the result will be very complex. But if the coupling strength is very small, only the subspace spanned by the lowest band need to be considered. Then the effective Hamiltonian can be written as a two-dimensional matrix using {these} two states as {bases}. The diagonal term is identical, the off-diagonal term is the coupling between two spin components which can be estimated as $\Omega_{k,\varphi}B(\varphi)$ where
\begin{equation}
\Omega_{k,\varphi}=\int^d_0\psi^*_{k\varphi,\downarrow}(x)\psi_{k\varphi,\uparrow}(x)dx=\sum_\ell(-1)^\ell|c_\ell(k,\varphi)|^2
\end{equation}
It can be validated numerically that $\Omega_{k,\varphi}$ is always positive in the first BZ. Then we need to discuss two different cases that $B(\varphi)$ is real and complex separately.

\subsection{The spatial and temporal domain are separable}

Assume that the coupling has the form $B(\varphi)=B_0(1+\cos\varphi)$, then the effective Hamiltonian can be written as
\begin{equation}
\mathcal{H}_\mathrm{eff}(k,\varphi)=\epsilon_{k,\varphi}\mathcal{I}+B_0(1+\cos\varphi)\Omega_{k,\varphi}\sigma_x
\end{equation}
The eigenstates of this effective Hamiltonian can be written as
\begin{equation}
|\psi^\pm_{k,\varphi}(x)\rangle=\{|\psi_{k\varphi,\uparrow}(x)\rangle\pm|\psi_{k\varphi,\downarrow}(x)\rangle\}/\sqrt{2}=\frac{e^{ikx}}{\sqrt{2d}}\sum_\ell c_\ell(k,\varphi)\exp\Big\{\frac{i2\pi\ell x}{d}\Big\}\big(|\uparrow\rangle\pm(-1)^\ell|\downarrow\rangle\big)
\label{groundstateseparable}
\end{equation}
The sign of the ground state just depends on the sign of coupling $B_0+B'\cos(\varphi)$. For example, if $B_0>B'>0$, it is always positive which means $|\psi^-_{k,\varphi}(x)\rangle$ is always the ground state, as shown in Fig. S4(a). The form of non-Abelian Berry connection can be estimated as
\begin{equation}
i\langle u^+_{k\varphi}(x)|\partial_\varphi|u^+_{k\varphi}(x)\rangle=i\langle u^-_{k\varphi}(x)|\partial_\varphi|u^-_{k\varphi}(x)\rangle=i\sum_\ell c^*_\ell(k,\varphi)\partial_\varphi c_\ell(k,\varphi),\,\,\,\,\,\,\,\,\,\,\,\,i\langle u^+_{k\varphi}(x)|\partial_\varphi|u^-_{k\varphi}(x)\rangle=0.
\end{equation}
which only has the diagonal term. From the Eq. (6) in main text, we know that the form of {the} Wilson line in this case is a straight line, there is no population transition in the pumping process.

\subsection{The spatial and temporal domain are {inseparable}}

When the coupling has the form $B(\varphi)=B_0(1+e^{i\varphi})$, the result will be very different. In this case, the coupling is complex so the effective Hamiltonian has a different form as
\begin{equation}
\mathcal{H}_\mathrm{eff}(k,\varphi)=\epsilon_{k,\varphi}\mathcal{I}+B_0(1+\cos\varphi)\Omega_{k,\varphi}\sigma_x-B_0\sin\varphi\Omega_{k,\varphi}\sigma_y
\end{equation}
Because the off-diagonal terms can be rewritten as $2B_0\cos(\varphi/2)e^{i\varphi/2}\Omega_{k,\varphi}$, which has an additional phase $e^{i\varphi/2}$, the eigenstates of the effective Hamiltonian are
\begin{equation}
|\psi^\pm_{k,\varphi}(x)\rangle=\{|\psi_{k\varphi,\uparrow}(x)\rangle\pm e^{-i\varphi/2}|\psi_{k\varphi,\downarrow}(x)\rangle\}/\sqrt{2}=\frac{e^{ikx}}{\sqrt{2d}}\sum_\ell c_\ell(k,\varphi)\exp\Big\{\frac{i2\pi\ell x}{d}\Big\}\big(|\uparrow\rangle\pm(-1)^\ell e^{-i\varphi/2}|\downarrow\rangle\big)
\label{groundstateunseparable}
\end{equation}
Then if the pumping process is in the temporal domain, the matrix elements of non-Abelian Berry connection is
\begin{equation}
\begin{split}
&i\langle u^+_{k\varphi}(x)|\partial_\varphi|u^+_{k\varphi}(x)\rangle=i\langle u^-_{k\varphi}(x)|\partial_\varphi|u^-_{k\varphi}(x)\rangle=i\sum_\ell c^*_\ell(k,\varphi)\partial_\varphi c_\ell(k,\varphi)+\frac{1}{4}, \\
&i\langle u^+_{k\varphi}(x)|\partial_\varphi|u^-_{k\varphi}(x)\rangle=i\langle u^-_{k\varphi}(x)|\partial_\varphi|u^+_{k\varphi}(x)\rangle=-\frac{1}{4}.
\end{split}
\end{equation}
Because $\alpha_k(\varphi)=i\sum_\ell c^*_\ell(k,\varphi)\partial_\varphi c_\ell(k,\varphi)$ is a real number, the form of Wilson line can be estimated as
 \begin{equation}
 \begin{split}
 \mathcal{W}_{0\rightarrow\varphi}=&\exp\Big\{\mathcal{T}\int^\varphi_0 i\boldsymbol{\mathcal{A}}(\varphi)d\varphi\Big\}=\exp\Big\{\mathcal{T}\int^\varphi_0 i\Big[\Big(\alpha_k+\frac{1}{4}\Big)\mathcal{I}-\frac{1}{4}\sigma_x\Big]d\varphi\Big\} \\
 =&e^{i\theta(\varphi)}\exp\Big\{-\frac{i}{4}\int^\varphi_0\sigma_x d\varphi\Big\}=e^{i\theta(\varphi)}\left(\begin{array}{cc} \cos(\varphi/4) & -i\sin(\varphi/4) \\ -i\sin(\varphi/4) & \cos(\varphi/4) \end{array}\right)
 \label{wilsonline}
 \end{split}
 \end{equation}
 where $\theta(\varphi)=\int^\varphi_0(\alpha_k+1/4)d\varphi$ is the global phase. If the initial state is prepared on the ground state $|\psi^-_{k,\varphi}(x)\rangle$, the populations of different bands in the pumping process should be
 \begin{equation}
 |W_{--}|^2=\cos^2\frac{\varphi}{4}=\frac{1}{2}\Big(1+\cos\frac{\varphi}{2}\Big)\,\,\,\,\,\,\,\,\,\,\,\,\,\,\,\,\,\, |W_{-+}|^2=\sin^2\frac{\varphi}{4}=\frac{1}{2}\Big(1-\cos\frac{\varphi}{2}\Big)
 \label{popwilsonline}
 \end{equation}
 which have a period of $4\pi$.
 Considering the glide operation along the synthetic dimension where $|\uparrow\rangle\rightarrow|\downarrow\rangle$ and $|\downarrow\rangle\rightarrow e^{i\varphi}|\uparrow\rangle$, $|\psi^\pm_{k,\varphi}(x)\rangle$ are the eigenstates of this synthetic glide operator with eigenvalues ``$+$" and ``$-$" respectively. But we should point out that the amplitude of coupling $2B_0\cos(\varphi/2)\Omega_{k,\varphi}$ will change the sign after passing the edge of the first BZ, it means that in the range of $\pi\le\varphi\le 3\pi$, $|\psi^+_{k,\varphi}(x)\rangle$ will be the ground state and the population on the ground state should changes from $|W_{--}|^2$ to $|W_{-+}|^2$. That is the reason we can get the result in Fig. 3(b) in the main text \cite{dssh}.

 If the pumping process is along the diagonal direction of the BZ of the effective Hamiltonian, we should redefine new parameters $p=(kd+\varphi)/2$ and $q=(kd-\varphi)/2$. The inverse transformation is $kd=p+q$ and $\varphi=p-q$. The eigenstates can be described using new coordinates as
\begin{equation}
|u^\pm_{pq}(x)\rangle=\frac{1}{\sqrt{2d}}\sum_\ell c_\ell(p,q)e^{2\pi i\ell x/d}\big\{|\uparrow\rangle\pm(-1)^\ell e^{i(q-p)/2}|\downarrow\rangle\big\}
\end{equation}
One can also estimate all elements of non-Abelian Berry connection along $p$ direction using the similar method and the form of Wilson line is the same as Eq. (\ref{wilsonline}). But in this case, the difference is that the eigenstates $|u_{pq,+}(x)\rangle$ and $|u_{pq,-}(x)\rangle$ will not flip when crossing the edge of the first BZ, since, although $2B_0\cos(\varphi/2)$ flips the sign, the quasi-momentum $k$ also goes through the edge of BZ, which results in $\Omega_{k,\varphi}=\sum_\ell(-1)^\ell|c_\ell(k,\varphi)|^2$ fliping the sign as well. This result comes from the properties of Bloch wave function that $c_{\ell}(k+2\pi/d)=c_{\ell+1}(k)$. So the coupling amplitude $2B_0\cos(\varphi/2)\Omega_{k,\varphi}$ will not change the sign and the population on the lowest band will always be $|W_{--}|^2$ as shown in Eq. (\ref{popwilsonline}), which leads to the result in Fig. 3(c) in the main text.

With the increase of the coupling amplitude $\Omega(\varphi)$, the influence of higher bands of Hamiltonian $\mathcal{H}_\sigma(x,\varphi)$ such as $p$ band can not be ignored. The lowest four bands need to be considered and there is no analytical {result} in this case. With numerical calculation, one can find that the {dominant} population on the lowest two bands should be $|\psi^{s,-}\rangle$ and $|\psi^{p,-}\rangle$ instead of $|\psi^{s,+}\rangle$ and $|\psi^{s,+}\rangle$ after passing the critical point of topological phase transition. But the qualitative properties of Wilson loop along two different routes still remain unchanged.

\end{appendix}

\end{document}